\documentclass[aps,prfluids,reprint,superscriptaddress,onecolumn,floatfix]{revtex4-2}
\usepackage{silence}
\usepackage[utf8]{inputenc}
\usepackage[english]{babel}
\usepackage{mathtools,amssymb,amsthm,relsize,thmtools,mathrsfs}
\usepackage{amsfonts}
\usepackage{graphicx,xcolor,color,colortbl}
\usepackage{soul,cancel}
\usepackage[shortcuts]{extdash}
\usepackage[colorlinks=true,linkcolor=blue,citecolor=blue,urlcolor=blue]{hyperref}

\makeatletter
\def\@bibdataout@rev{%
 \immediate\write\@bibdataout{%
  @CONTROL{%
   REVTEX42Control,
   \eprint@enable@sw{}{,eprint="1"}%
  }%
 }%
 \if@filesw
  \immediate\write\@bibdataout{%
   @CONTROL{%
    apsrev42Control%
    \longbibliography@sw{%
     ,author="08",editor="1",pages="0",title="0",year="1"%
    }{}%
   }%
  }%
 \fi
}%
\makeatother

\renewcommand{\vec}[1]{\boldsymbol{#1}}

\newcommand{\Rey}{\mathit{Re}}
\newcommand{\Ros}{\mathit{Ro}}
\renewcommand{\d}{\mathrm{d}}

\newcommand{\note}[1]{\textcolor{blue}{#1}}
\newcommand{\error}[2]{\ifmmode\textcolor{red}{\cancel{#1}}\ifstrempty{#2}{}{\note{\to#2}}\else\textcolor{red}{\st{#1}}\ifstrempty{#2}{}{\note{\textrightarrow#2}}\fi}

\begin{document}

\title{Topographic Effects on Steady-States of Non-Rotating Shallow Flows}

\author{Pierpaolo Bilotto}
\email{pierpaolo.bilotto@gssi.it}
\affiliation{\foreignlanguage{italian}{Gran Sasso} Science Institute, \foreignlanguage{italian}{L'Aquila}, Italy}

\author{Roberto Verzicco}
\affiliation{\foreignlanguage{italian}{Gran Sasso} Science Institute, \foreignlanguage{italian}{L'Aquila}, Italy}
\affiliation{PoF, University of Twente, Enschede, The Netherlands}

\date{\today}

\begin{abstract}
	\noindent
	In this work, we discuss the long\-/time behavior of non\-/rotating quasi-2D viscous flows over topographies.
	We develop a novel theoretical and numerical framework for the analysis of these flows, derived as a dimensional reduction of the 3D Navier\-/Stokes equations in the limit of infinite Rossby number \(\Ros\).
	We numerically determine dynamical attractors for fixed kinetic energy, focusing on the dependence of the final state on the Reynolds number.
	Under turbulent conditions, the attractor is no longer unique but delocalized, spanning the lowest excited states of the deterministic system.
	Regardless of the realized stationary configuration, large\-/scale vortices settle within topographic valleys, in contrast with the phenomenology of the rotating case.
	These findings have significant implications for understanding steady turbulent regimes in slowly rotating (\(\Ros\gg1\)) planetary environments.
\end{abstract}

\maketitle

\section{Introduction}
	\label{sec:intro}
	In geophysical contexts, pseudo-stationary features characterize most of the oceanic and atmospheric currents responsible for major climate events.
	These large\-/scale flows are typically confined in shallow domains where the motion is predominantly two\-/dimensional \cite{alexakis23}.
	In these conditions, 2D turbulence describes the emergence of long\-/lived coherent structures through an inverse energy cascade, as explained in \cite{kraichnan80, boffetta12}.
	The stability and the relaxation toward these 2D steady states have been widely investigated employing tools from nonlinear stability theory and Miller\-/Robert\-/Sommeria theory \cite{arnold65,pierini81,carnevale87,miller90,robert91}.
	One of the most intriguing aspects of the theory is the emergence of domain-sized vorticity condensates \cite{kraichnan67,smith94,laurie14}.

	This rich phenomenology is shared by several geophysical models developed as specializations of the Navier-Stokes equations.
	Most of them also share the assumption of small Rossby numbers \(\Ros=U/fL\), where \(U\) is the typical velocity scale, \(f=2\Omega\sin\phi\) the Coriolis parameter at latitude \(\phi\), and \(L\) the relevant horizontal length scale of the flow.
	On fast rotating systems (\(\Ros\ll1\)), the strong Coriolis force dominates over inertia and arranges the flow into Taylor columns, aligned with the rotation axis, effectively linearizing the interaction of the flow with the underlying topography.
	Within this broadly\-/applicable and simplified framework, extensive research has characterized geostrophic flows over topography \cite{vidal24,siegelman23, gallet24}: these studies explain jet formation and flow\-/topography alignment, while also providing insights into the nature of equilibria and energy transport in rotating turbulent systems \cite{bretherton76,herring77,vallis93,majda97,renaud16}.

	However, strong rotation is not ubiquitous in geophysical flows: examples range from the super\-/rotating atmosphere of slowly spinning bodies, like Venus (\(\Ros\sim370\)) or Titan (\(\Ros\sim18\)) \cite{read11,read18}, to the dynamics of Earth's equatorial oceans (\(\Ros\to\infty\)).
	When the Coriolis force is negligible, the topographic interaction is fully nonlinear.
	A rigorous derivation of an effective 2D approximation is effectively achieved by using a mass\-/transport stream function to account for the variable fluid depth, as advocated by \cite{evans92}.
	While the resulting equations are analytically and numerically more challenging, this less restrictive model relaxes the usual geophysical assumption of small topography, as recently explored by \cite{maranzoni22,maranzoni23}.
	Overall, the nonlinear topographic coupling creates a richer dynamical scenario where the system may explore multiple metastable states instead of settling into its ground state, similarly to what already discussed in \cite{bouchet12}.

	In this work, we present numerical results on the stability of quasi\-/2D viscous flows over topography in the absence of rotation.
	We focus on their long\-/time behavior, determining whether such a system can achieve stationarity, either for fixed kinetic energy or driven by a stochastic forcing. 
	Our results show that, regardless of the conditions, vortices consistently avoid topographic hills, in contrast with the predictions for the rotating case.
	While for small turbulence Reynolds number, \(\Rey_L=\sqrt{E}/\nu\), the flow directly reaches its ground state, for sufficiently high kinetic energy \(E\) and low kinematic viscosity \(\nu\), the dynamics can be temporally trapped in excited states.
	Finally, the random forcing prevents turbulent flows from relaxing to a unique ground state.
	These findings have significant implications for understanding out\-/of\-/equilibrium turbulent regimes in planetary environments where rotation is weak or absent.

	Our non\-/rotating shallow flow model (NRSF) is derived from the 3D incompressible Navier-Stokes equations in Sec.~\ref{sec:model}.
	In Sec.~\ref{sec:numerics}, we describe a novel approach to the numerical integration of the NRSF, that handles the complication of dealing with a mass\-/transport stream function.
	Sec.~\ref{sec:results} presents the results of our numerical experiments, showing the long\-/time behavior of the NRSF for different Reynolds numbers and with two kinds of external energy inputs, one deterministic and the other stochastic.
	Finally, Sec.~\ref{sec:conclusion} provides a summary of our findings and points at directions for future work.

\section{Model}
	\label{sec:model}
\begin{figure}[b]
		\centering
		\includegraphics[width=0.5\textwidth]{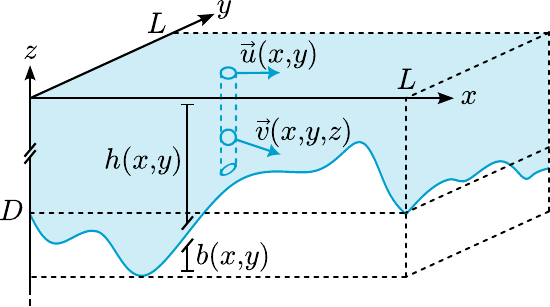}
		\caption{An illustration of the fluid with the relevant variables}.
		\label{fig:sketch}
	\end{figure}

	We are interested in studying the long time behavior of a laterally unbounded three\-/dimensional fluid layer confined in the vertical direction by a stress\-/free bottom topography and a top flat lid.
	For computational reasons, we will only consider periodic topographies and, therefore, model the infinite domain as a periodic square.
	Hence, the 3D domain is \(\mathscr{D} = \mathbb{T}\times[-h,0]\), where \(h(x,y)>0\) is the local height of the fluid column (see Fig.~\ref{fig:sketch}).
	We assume the 3D flow to be incompressible with homogeneous density \(\rho = 1\) and to evolve according to the Navier\-/Stokes equations,
	\begin{align}
		\label{eqn:NS}
		&D_t\vec{v} = -\nabla{p} + \nu\Delta\vec{v} + \vec{\mathrm{f}},\\
		\label{eqn:3D_incompressibility}
		&\nabla\cdot\vec{v} = 0,
	\end{align}
	where \(D_t:=\partial_t+(\vec{v}\cdot\nabla)\) the material derivative, \(\vec{v}(x,y,z)\) the 3D velocity, \(p(x,y,z)\) the pressure, \(\nu\) the kinematic viscosity, \(\Delta:=|\nabla|^2\) the Laplacian operator, and \(\vec{\mathrm{f}}(x,y,z)\) a generic external forcing.
	In a very shallow domain, where the vertical depth, \(D \simeq \langle{h}\rangle\), is much smaller than the horizontal size, \(L = 2\pi\), the fluid motion is approximately columnar, and the flow can be described by a class of models, generically called Shallow Water\,\cite{salmon_book,vallis_book}.
	More in detail, the assumption \(D \ll L\) implies that vertical velocity and acceleration are much smaller than the horizontal ones, and that the fluid is in hydrostatic equilibrium, that is, the pressure compensates for any vertical volume forcing\cite{footnote:pressure}.
	In this approximation, the horizontal equations are effectively decoupled from the vertical one.
	Thus, if we consider an initial datum and a forcing which are both independent of \(z\), the same will be true for the horizontal velocity \cite{footnote:boundary_terms}.
	In the end, the behavior of a 3\-/dimensional fluid in a very narrow domain is effectively described by a 2\-/dimensional system of equations on the torus,
	\begin{align}
		\label{eqn:SW_velocity}
		&\bar{D}_{t}\vec{u} = -\bar{\nabla}p + \nu\bar{\Delta}\vec{u} + \bar{\vec{\mathrm{f}}},
	\end{align}
	where we have introduced the 2D horizontal velocity \(\vec{u}\), the forcing \(\bar{\vec{\mathrm{f}}} = (\mathrm{f}_x,\mathrm{f}_y)\), and the horizontal gradient \(\bar{\nabla}\).
	(In the rest of this paper we will drop the over\-/bar for the 2D case, unless it causes ambiguity.)
	Within the same assumption, integrating Eq.~\eqref{eqn:3D_incompressibility} along the vertical direction gives \(-v_z\big{|}_b + h\,\nabla\cdot\vec{u} = 0\).
	Since the position at the fluid bottom is just \(z = -h\), we obtain a dynamical condition for the height of the fluid column,
	\begin{align}
		\label{eqn:SW_density}
		D_t{h} = -h\,\nabla\cdot\vec{u}.
	\end{align}
	
	Our system further simplifies by considering the scalar relative vorticity, \(\zeta = \nabla\times\vec{u}\),	where we used the curl notation with a 2D vector to denote the operator returning the vertical component of the 3D curl, \(\partial_x{u_y} - \partial_y{u_x}\).
	We take the curl of Eq.~\eqref{eqn:SW_velocity}, obtaining
	\[D_t\zeta + \zeta\,\nabla\cdot\vec{u} = \nu\Delta\zeta + \nabla\times\vec{\mathrm{f}},\]
	which can be combined with Eq.~\eqref{eqn:SW_density} to give
	\begin{align}
		\label{eqn:SW_unfinalized}
		D_t{q} = \frac{\nu}{h}\Delta\zeta + \frac{\mathfrak{f}}{h},
	\end{align}
	where \(q = \zeta/h\) is the potential vorticity, and \(\mathfrak{f}=\nabla\times\vec{\mathrm{f}}\).
	The convenience of the vorticity formulation lies in reducing the system to a single prognostic equation for the scalar field \(q\).
	This is particularly useful for an incompressible flow, where the velocity can be fully recovered from the vorticity, effectively reducing the system to only one unknown.
	In the following we will discard surface modulations in favor of a rigid lid, \(\partial_t{h} = 0\), which eliminates gravity wave dynamics and Froude number dependence.
	From the mass conservation, Eq.~\eqref{eqn:SW_density}, we can see that the mass-transport velocity (or momentum) \(h\vec{u}\) is divergence\-/less, and can be expressed in terms of a scalar field, \(\psi\), namely the stream function \cite{footnote:stream_function},
	\begin{align}
		\label{eqn:stream_function}
		h\vec{u} = \nabla^\perp\psi,
	\end{align}
	where we used the perpendicular gradient, \(\nabla^\perp := \mathbb{J}\nabla = (\partial_y,\,-\partial_x)\), with \(\mathbb{J}\) being the unitary skew\-/symmetric matrix with \(\mathbb{J}_{1,2} = 1\).
	The potential vorticity can be expressed in terms of the stream function as
	\begin{align}
		\label{eqn:potential_vorticity}
		q = \frac{1}{h} \nabla\times\left(\frac{\nabla^\perp\psi}{h}\right) = \mathcal{L}[\psi],
	\end{align}
	which can be inverted to obtain an implicit expression for \(\psi[q] = \mathcal{L}^{-1}[q]\), where
	\begin{align}
		\label{eqn:PV_SF}
		\mathcal{L} := \frac{1}{h^2} \left[-\Delta + \frac{\nabla{h}}{h}\cdot\nabla\right].
	\end{align}
	The equation of motion for \(q\) can now be closed by expressing \(\vec{u}\) in terms of \(\psi\) (that, in turn, depends on \(q\)), leading to the final equation object of our study:
	\begin{align}
		\label{eqn:NRSF_equation}
		h\,\partial_t{q} + J(q,\psi) = \nu\Delta(hq) + \mathfrak{f},
	\end{align}
	where we introduced the Jacobian notation \(J(f, g) := \det(\nabla{f},\nabla{g}) = \nabla{f}^T\,\mathbb{J}\,\nabla{g}\).
	Eq.~\eqref{eqn:NRSF_equation} and \eqref{eqn:potential_vorticity} constitute the non\-/rotating shallow flow model (NRSF).

	As a consequence of the rigid lid assumption, the height of the fluid column is only fixed by the choice of bottom topography, denoted by \(b(x,y)\).
	Without loss of generality, we assume \(b\in[-1,1]\).
	We will employ a recipe that always leaves some space for the fluid, that is \(h=1-\alpha\,b>0\), for \(\alpha\in(0,1)\).
	In the numerical experiments presented hereafter, \(\alpha\) was set to \(0.1\).
	The simplest topographies considered here reproduce an isolated hill or valley, shaped as a periodic Gaussian with \(\sigma = L/10\) placed at the center of the domain.
	To mimic a generic Earth\-/like landscape, instead, we build topographies as sums of Gaussian functions with random height, width, sign and center position.

\section{Numerical Method}
	\label{sec:numerics}
	To test our hypotheses throughout the text, we performed numerical simulations of Eq.~\eqref{eqn:NRSF_equation} in different configurations.
	In this section, we give an overview of the employed numerical recipe, specifically tailored to handle the formulation in terms of the mass-transport stream function.

	We numerically solve Eq.~\eqref{eqn:NRSF_equation} in a periodic square of size \(L = 2\pi\) using a second\-/order finite difference method on an \(N_x \times N_y\) grid.
	All the simulations presented in this paper are performed with \(N_x\times N_y=512^2\) grid points.
	Two integration schemes are employed in the time evolution: a Crank\-/Nicolson implicit scheme, taking care of the linear terms, is embedded into a third\-/order explicit Runge\-/Kutta scheme for nonlinear and external terms, as detailed in \cite{orlandi_book}.

	The numerical scheme to compute each of the terms must enforce the conservation of the symmetries, as discussed in Sec.~\ref{sec:symmetries}.
	The adopted five\-/point stencil scheme for the Laplacian \cite{abramowitz65}, like its continuous counterpart, preserves all the domain symmetries and anti-symmetries.
	The nonlinear transport term \(J(q,\psi)\) is computed employing Arakawa's second order scheme\,\cite{arakawa66}, which ensures the conservation of energy, enstrophy and the skew symmetry of the Jacobian under exchange of \(q\) and \(\psi\).
	In addition, it can be shown that this scheme shares the same symmetry properties as the continuous Jacobian (refer to Sec.~\ref{sec:symmetries}).

	The novelty of the approach, with respect to the usual scheme for the Euler equation, lies in the use of a stream function for the incompressible mass-transport velocity, \(h\vec{u}\), instead of the usual velocity\,\cite{fletcher91}.
	This entails a more complicated expression for \(\psi_t\), as the operator \(\mathcal{L}\) in Eq.~\eqref{eqn:PV_SF} cannot be efficiently inverted in Fourier space.
	We resort to a recursive approach.
	At each time step, we initialize an auxiliary variable, \(\Psi_t^{n}\), as
	\[\Psi_t^0=\psi_{t-1}.\]
	The next step involves updating \(\Psi_t\) according to the rule
	\[\Psi_t^{n+1} = \Delta ^{-1}\left[\frac{\nabla{h}}{h}\cdot\nabla\Psi_t^n - h^2q_t\right],\]
	where the Laplacian is inverted in Fourier space.
	By construction, the fixed point of this discrete process, \(\Psi_t^\infty\), satisfies Eq.~\eqref{eqn:potential_vorticity} and, therefore, represents the updated stream function, \(\psi_t\).
	Numerically, the iteration proceeds until convergence, that is	\(\max|\Psi_t^n-\Psi_t^{n-1}| < 10^{-15}\,\max|\Psi_t^n|\).
	With our choice of initialization, the recursion usually converges in no more than 10 steps.
	Computing gradients of \(\psi\) and \(h\) along the coordinate axes breaks the symmetries preserved by the continuous equation \eqref{eqn:NRSF_equation} (see Appendix~\ref{sec:symmetries}).
	Since these gradients only appear in a scalar product, which is rotationally invariant, to ensure that \(\psi\) inherits the symmetries properties of \(q\), we can compute them along diagonal grid points,
	\[(\nabla{f}\cdot\nabla{g})_{i,j} = (f_{i+1,j+1} - f_{i-1,j-1})(g_{i+1,j+1} - g_{i-1,j-1}) + (f_{i+1,j-1} - f_{i-1,j+1})(g_{i+1,j-1} - g_{i-1,j+1}),\]
	rather than along the coordinated horizontal axes.

	We performed simulations both with and without an external forcing.
	The velocity forcing employed, \(\vec{\mathrm{f}}_r\), is random but spatially homogeneous.
	Its spectrum is extracted in such a way to be uniformly distributed in wave numbers \(k\) between \(k_r-\Delta{k}_r/2\) and \(k_r+\Delta{k}_r/2\), with \(k_r = 50\) and \(\Delta{k}_r = 3\), and zero elsewhere.
	The forcing amplitude matrix is fixed in order to supply a constant power, \(\epsilon = 1/2\,\int_{\mathbb{R}^+}\langle\vec{\mathrm{f}}_r(t)\cdot\vec{\mathrm{f}}_r(t+s)\rangle\d{s}\), in stationary conditions.
	Practically, the vorticity forcing \(\mathfrak{f}_r\) is computed directly as the anti\-/transform of the curl spectrum, as explained in \cite{alvelius99}.

	\subsection{Validation of the Numerical Method}
		\label{sec:numerical_validation}
		\begin{figure*}[t]
			\centering
			\includegraphics[width=\textwidth]{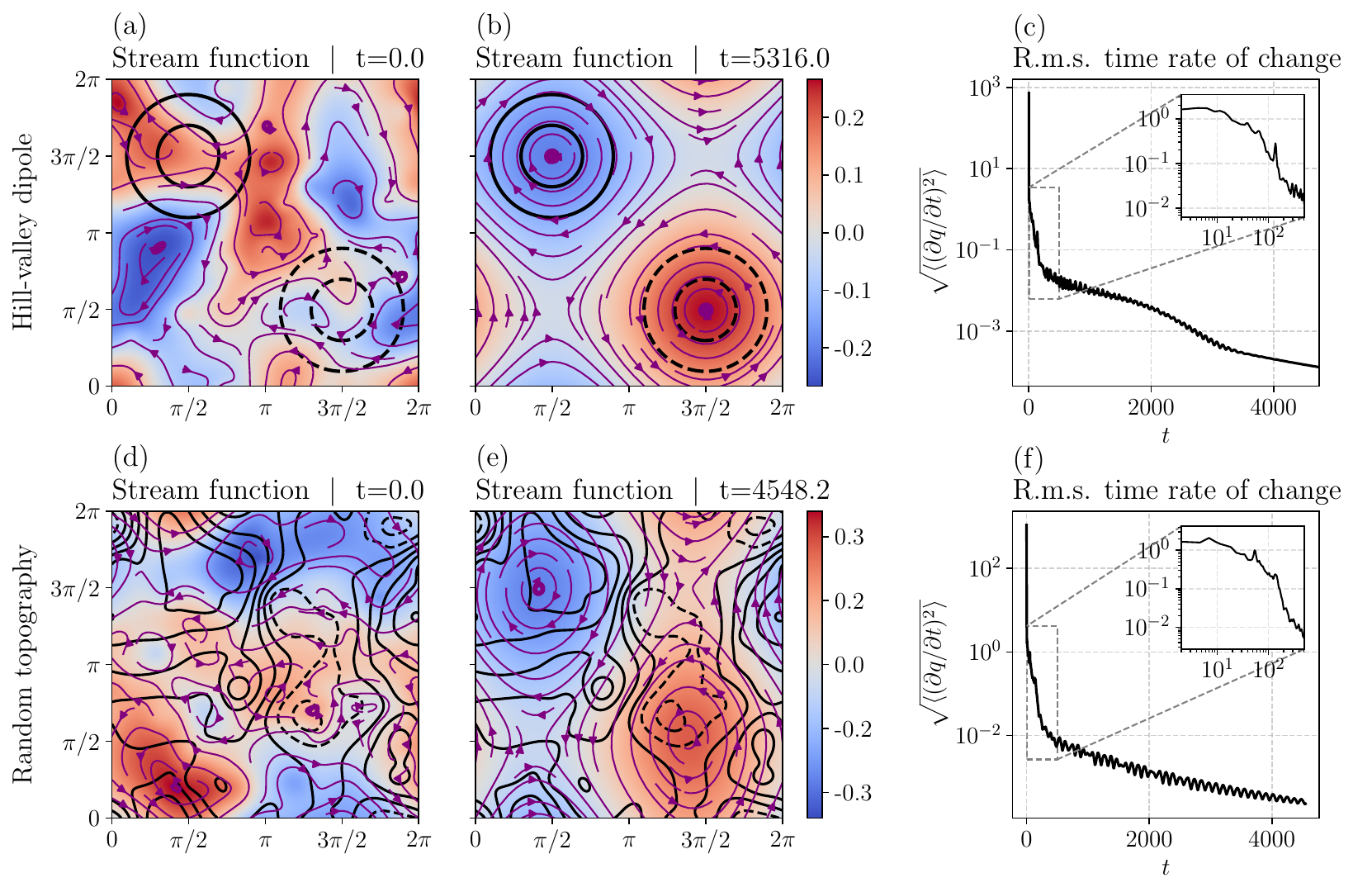}
			\caption{
				Plots of the stream function (a, b, d, e) and of the rate of change (c, f) for two simulations in a rotating frame, as detailed in Sec.~\ref{sec:numerical_validation}.
				In the top row, the topography is a dipole of a Gaussian hill and valley, while the bottom row employs a random topography.
				In panels (a, b, d, e), the colors show the stream function, black lines represent topographic contours (negative values are dashed, black lines represent the contours at one and two \(\sigma\) of the Gaussian), and purple arrows point along streamlines of the velocity field.
				The insets in panels (c, f) zoom on the transient during which coherent vortices form.
			}
			\label{fig:rot_sf}
		\end{figure*}
		To validate this novel approach, we benchmark it against one of the first works regarding flows above topography.
		In section 8 of their paper \cite{bretherton76}, Bretherton and Haidvogel observed the relaxation of a random initial condition in a rotating reference frame on a random shallow topography.
		Eq.~\eqref{eqn:NS} can be easily adapted to a rotating frame by adding a body force in the form of the Coriolis term \(-\vec{\mathrm{f}}\times\vec{v}\).
		By repeating the same derivation as in Sec.~\ref{sec:model}, the transported quantity in Eq.~\eqref{eqn:SW_unfinalized} is
		\begin{align}
			\label{eqn:rotating_PV}
			q=\frac{\zeta+f}{h}
		\end{align}
		where \(f=\Omega\cos(\varphi)\) is the scalar Coriolis factor at latitude \(\varphi\).
		For our validation purpose, we will consider a constant Coriolis factor \(f=1\).

		We run two simulations on different topographies, \(b(x,y)\): the first is a dipole of a Gaussian hill and valley, maximally spaced on the torus, while the second is obtained as the sum of 100 random hills and valleys.
		In both cases, the initial state \(\psi_0\) (and accordingly \(q_0\)) is random and uncorrelated with the topography, normalized to have unitary energy.
		Both the initial stream function field and the topography are shown in panels (a) and (d) of Fig.~\ref{fig:rot_sf}.
		In panel (c) and (f) of the same figure, one can distinguish two regimes, representing two different phases of the time evolution.
		During the first transient period (approximately until \(t=200\)) the state moves away from the random initial condition to a more topographically favorable state, forming two opposing large\-/scale vortices.
		Later, the clockwise and anti-clockwise vortices slowly settle onto the respective valley and hilltop, as expected in the Northern Hemisphere, where \(f>0\).
		Their oscillation around the stable position is mirrored by the oscillation of the time derivative root\-/mean\-/square.
		The final snapshot of the dipole topography, in panel (b) of Fig.~\ref{fig:rot_sf}, shows the vortices aligned to the topography contours, in agreement with the theory of rotating geostrophic flows.
		On a random topography the final alignment is not as precise, as seen in panel (e).
		Vortices still settle in the extremes of the topography, but their contours are coarse-grained by the viscous dissipation.
		A more accurate alignment could be accomplished by using a smaller scale dissipation, like the hyperviscosity employed in the original paper \cite{bretherton76}.

		Finally, we remark that the observed behavior is indeed consistent with the expectation in a rotating geophysical system close to equilibrium, but is not the true long\-/time final state of the simulated dissipative system.
		Expanding the expression of \(q\) in powers of the small topography correction \(\alpha\), \(q\simeq f\,(1+\alpha b)+\zeta\), one can see how the potential vorticity, and not the stream function, aligns to the topography \(b\) when viscosity eventually dissipates most of the relative vorticity \(\zeta\).
		At the same time, the stream function may condense at middle altitudes, as the flow is too weak to squeeze over hills and steers around them.

\section{Results and Analysis}
	\label{sec:results}
	In our investigation of quasi\-/steady states, the dissipative nature of the system raises an issue.
	On one hand, dealing with an unforced dissipative system, one may have to resort to less strict definitions of stationarity, appropriate for a decaying state.
	On the other hand, an external forcing would input the necessary energy to establish a forced\-/dissipative steady state; the drawback being that the system would be externally driven towards a specific forcing\-/dependent state.
	In the following, we delve into both alternatives, addressing the problems of decaying flows, discussing their solutions, and trying to generalize our findings to a more realistic externally forced system.

	An unforced viscous system, like the one described by Eq.~\eqref{eqn:NRSF_equation} for \(\vec{\mathrm{f}} = 0\), cannot preserve energy.
	The expression of the kinetic energy in our NRSF model derives from that of its original 3D counterpart, \(\mathscr{E}[\vec{v}] = 1/2 \int_{\mathscr{D}}|\vec{v}|^2\,\d^3{x} \simeq 1/2 \int_{\mathbb{T}} h|\vec{u}|^2\,\d^2{x}\); using the definition of stream function and integrating by parts results in
	\[\mathscr{E}[q] = \frac{1}{2}\int_\mathbb{T}\psi\,hq\,\d\vec{x}.\]
	The energy rate of change \(\partial_t\mathscr{E}\) is readily obtained by multiplying Eq.~\eqref{eqn:NRSF_equation} by \(\psi\) and integrating over the domain.
	One can see that the Jacobian term vanishes after an integration by parts, and what remains is the diffusive term
	\[\partial_t\mathscr{E} = \nu\int_\mathbb{T}\Delta{\psi}\,hq\,\d\vec{x}
	= \nu\int_\mathbb{T}\frac{1}{h}\Delta{\psi}\left[-\Delta{\psi} + \frac{\nabla{h}}{h}\cdot\nabla{\psi}\right]\d\vec{x},\]
	which is composed of the negative definite dissipative term, in common with the flat bottom case (2D Navier\-/Stokes equation), and a boundary term.
	Despite the latter not being negative definite, energy decreases on average \cite{footnote:boundary_terms}, and we thus still refer to the system as dissipative.

	Under these conditions, regular stationarity is not admissible.
	In an attempt to address the decaying nature of the system, one could define a weaker notion of stationarity, only concerning the shape of the state during the viscous dynamics,
	\[q(\vec{x},t) = \epsilon(t)\,q_s(\vec{x}),\]
	where \(q_s\) is the stationary shape of the solution, and the coefficient \(\epsilon(t)\to0\) as energy is being dissipated.
	Such a state would satisfy the equation \(\partial_t{q} = (\dot{\epsilon}/\epsilon)\,q\).
	Substituting this stationarity ansatz into Eq.~\eqref{eqn:NRSF_equation}, we obtain
	\begin{align}
		\label{eqn:scale_not_stationary}
		(\dot{\epsilon}/\epsilon)\,hq_s + \epsilon\,J(q_s,\psi_s) = \nu\Delta(hq_s),
	\end{align}
	where \(\psi_s = \mathcal{L}^{-1}[q_s]\).
	For Eq.~\eqref{eqn:scale_not_stationary} to be satisfied for every value \(\epsilon(t)\) assumes during the dissipative dynamics, \(J\) must be identically zero \cite{footnote:jacobian} precisely when \(hq_s\) is an eigenvector of the Laplacian.
	For a generic non-uniform topography, these two independent constraints form an over\-/determined system that admits no non\-/trivial solution, unlike the flat\-/bottom case where \(h\) is constant \cite{foias84}.

	All the considerations so far lead to the conclusion that the dissipative NRSF equation cannot attain stationarity unless its scale is held constant.
	One easy and effective way to achieve this is by preserving the energy during the evolution.
	This can be achieved through the addition of an ad\-/hoc forcing that perfectly balances the dissipated energy, in the shape of \(-g\,\partial_t\mathscr{E}/\int g\,\psi\,\d{\vec{x}}\), for any function \(g(\vec{x})\neq0\).
	The natural choice for \(g\), not steering the system towards any predetermined state, is \(g = hq\), which leads to the non\-/local autonomous equation
	\begin{align}
		\label{eqn:fixed_energy}
		h\,\partial_t{q} + J(q,\psi) = \nu\Delta(hq) - \nu\frac{\int_\mathbb{T}\Delta{\psi}\,hq\,\d{\vec{x}}}{2\,\mathscr{E}[q]}\,hq,
	\end{align}
	which fixes the energy to the initial value \(E=\mathscr{E}[q_0]\), selecting attractors on a constant\-/energy manifold.
	This constraint converts the problem from a free\-/decaying flow into forced dissipative turbulence.
	Now, rescaling the solution as \(q = \sqrt{E}\,\tilde{q}\), we can look for stationary states as the long\-/time solutions to the equation
	\begin{align}
		\label{eqn:stationary_energy}
		h\,\partial_\tau\tilde{q} = -J(\tilde{q},\tilde{\psi}) + \frac{1}{\Rey_L}\left[\Delta(h\tilde{q}) - \frac{h\tilde{q}}{2}\int_\mathbb{T}\Delta{\tilde{\psi}}\,h\tilde{q}\,\d{\vec{x}} \right],
	\end{align}
	where \(\tau = \sqrt{E}\,t\) is an ``advective'' time\-/scale and \(\Rey_L = \sqrt{E}/\nu = \langle{u}\rangle L/\nu\) is the turbulence Reynolds number, where we used angle brackets to represent the root\-/mean\-/square value.
	We will refer to the expression in square brackets as the ``diffusive\-/conservative'' term (\(DC\)).
	Looking at Eq.~\eqref{eqn:stationary_energy}, it is clear that the stationary equation \(\partial_\tau\tilde{q}=0\) has a one-parameter family of solutions \(\tilde{q}_\infty(\vec{x};\Rey_L)\) depending on the value of \(\Rey_L\), which balances advection and diffusion.
	
	\begin{figure*}[b]
		\raggedright
		\includegraphics[width=\textwidth]{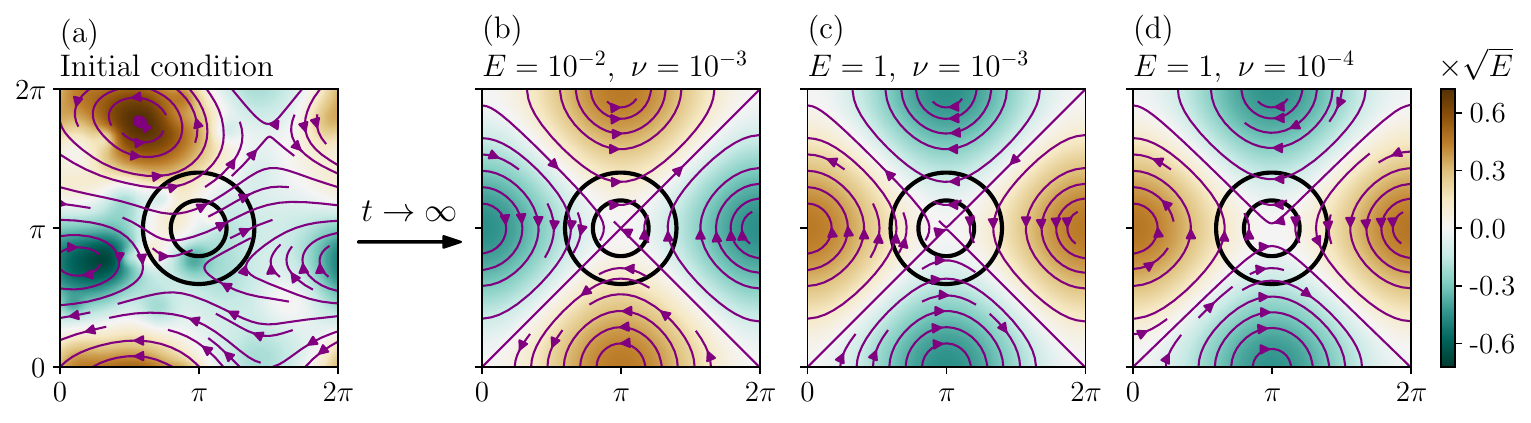}
		\caption{
			The initial (a) and final (b, c, d) state over a Gaussian hill, for three different configurations of parameters \(E\) and \(\nu\).
			Colors represent the rescaled potential vorticity \(\tilde{q}\), purple arrows point along streamlines of the velocity field, and black lines mark the contours at one and two \(\sigma\) of the Gaussian hill.
		}
		\label{fig:steadE_pv}
	\end{figure*}
	We ran numerical simulations of Eq.~\eqref{eqn:fixed_energy} starting from random initial conditions \(q_0\).
	The energy conservation is numerically implemented as a simple rescaling of the field, \(q_t\to\sqrt{E/\mathscr{E}[q_t]}\,q_t\), at each time step.
	We modeled the terrain as a simple Gaussian hill to isolate core topographic effects, while avoiding the multi\-/scale irregularities inherent in natural terrain.
	Here we present three simulations sharing the same rescaled initial condition, \(\sqrt{E}\,\tilde{q}_0\), represented in panel (a) of Fig.~\ref{fig:steadE_pv}.
	We examine three different parameter sets, (i) \(E=10^{-2}\) and \(\nu=10^{-3}\), (ii) \(E=1\) and \(\nu=10^{-3}\), and (iii) \(E=1\) and \(\nu=10^{-4}\), selected to span a representative range of Reynolds numbers, respectively \(Re_L=10^2\), \(10^3\), and \(10^4\).
	The final states are given in panels (b), (c), and (d) of Fig.~\ref{fig:steadE_pv}.

	\begin{figure*}[b]
		\centering
		\includegraphics[width=0.6\textwidth]{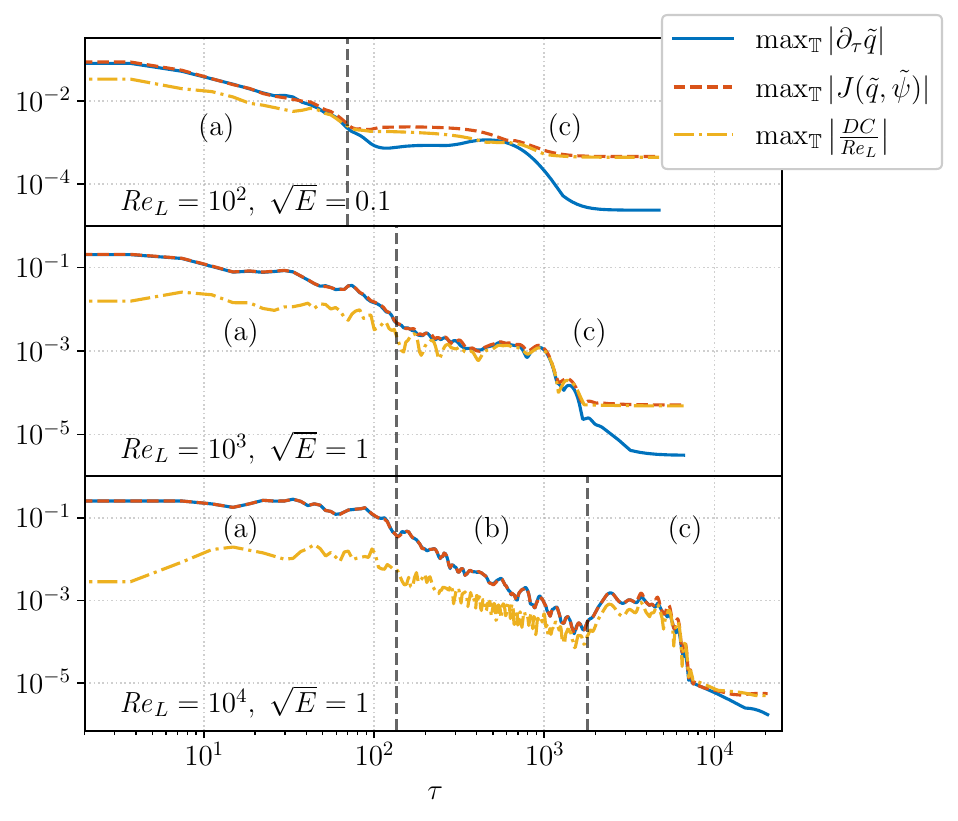}
		\caption{
			The maximum values of the terms in Eq.~\eqref{eqn:stationary_energy} as a function of the advective time \(\tau\), for the simulations shown in Fig.~\ref{fig:steadE_pv}.
			Different dynamical regimes are labeled as: (a), vortex dipole formation; (b), relaxation to a metastable state; (c), final relaxation to the fixed point.
		}
		\label{fig:steadE_dynamics}
	\end{figure*}
	All three simulations show the flow rapid development from the random initial condition into a large\-/scale vortex dipole.
	In runs (i) and (ii), this dipole slowly and coherently migrates towards its final position with different timescales, in agreement with the predicted scaling \(\tau=\sqrt{E}\,t\).
	Run (iii), however, exhibits a transient relaxation towards an intermediate state, characterized by its anti\-/symmetry about the diagonal axis.
	The flow remains suspended in this metastable state for several hundreds of time units, before eventually relaxing to the displayed final pattern.
	The distinct dynamical regimes are highlighted in Fig.~\ref{fig:steadE_dynamics}.

	In all investigated cases, by the end of the simulation, the time derivative of the potential vorticity drops (see Fig.~\ref{fig:steadE_dynamics}), indicating that the system has reached a numerically stationary state \cite{footnote:stability_test}.
	Even more interestingly, while for most of the dynamics the time derivative is dominated by the Jacobian term, the onset of stationarity is marked by the convergence of the advective and diffusive\-/conservative terms.
	Notably, in the late stages of the evolution their values remain appreciably larger than the time derivative, given by their difference.
	The persistence of this cancellation excludes the possibility of a non\-/advective stationary state, motivated by the minimum enstrophy (or selective decay) principle \cite{bretherton76,majda00}.
	This principle can be applied to several models of geophysical interest, the key assumption being that, under strong rotation \(f\gg1\) and with small topography \(\alpha\ll1\), Eq.~\eqref{eqn:rotating_PV} can be expanded at leading order around the average height \(\langle{h}\rangle=1\) as \(q=\zeta+f+\alpha f\,b\), where \(\alpha\) and \(f\) roughly compensate.
	The linear dependence on the topography \(b\) simplifies the treatment of the equation and enables the application of the minimum enstrophy principle, according to which the stationary state minimizes enstrophy at fixed energy.
	This condition implies that stream function and potential vorticity align, canceling the Jacobian term \cite{footnote:jacobian}.
	However, in the absence of a strong rotation, the nonlinear coupling between the flow and the topography prevents the system from being forced into a unique global attractor.
	In contrast, the cancellation of the advective and diffusive\-/conserving terms implies that the stationary solution \(\tilde{q}_\infty\) does depend on \(\Rey_L\), as the Reynolds number controls the relative magnitude of the two cancelling terms.
	This result diverges from the canonical predictions of the theory of selective decay, which proves that the stationary states do not depend on viscosity \cite{majda97}.

	Regardless of the Reynolds number, the final flow configuration in all three simulations is visually identical, differing only by a global sign inversion, a manifestation of spontaneous symmetry breaking in this chaotic system.
	The graphs show a couple of opposing vortices maximally spaced inside the periodic domain, as expected from the equilibrium configuration of a repulsive Ewald potential \cite{matthaeus91}.
	Remarkably, the dipole is located in such a way that both vortices are placed in the furthest possible positions from the hill.
	This is in stark contrast to the rotating case explored in Sec.~\ref{sec:numerical_validation}, where vorticity aligned to the topography, i.e. one vortex is located on top of the hill.
	This result can be intuitively explained invoking the conservation of potential vorticity \(q=\zeta/h\) along streamlines: as a fluid column moves up a hill, it gets vertically compressed, leading to a decrease in relative vorticity; conversely, when \(h\) stretches in a valley, \(\zeta\) increases accordingly.

	\begin{figure*}[t]
		\centering
		\includegraphics[width=\textwidth]{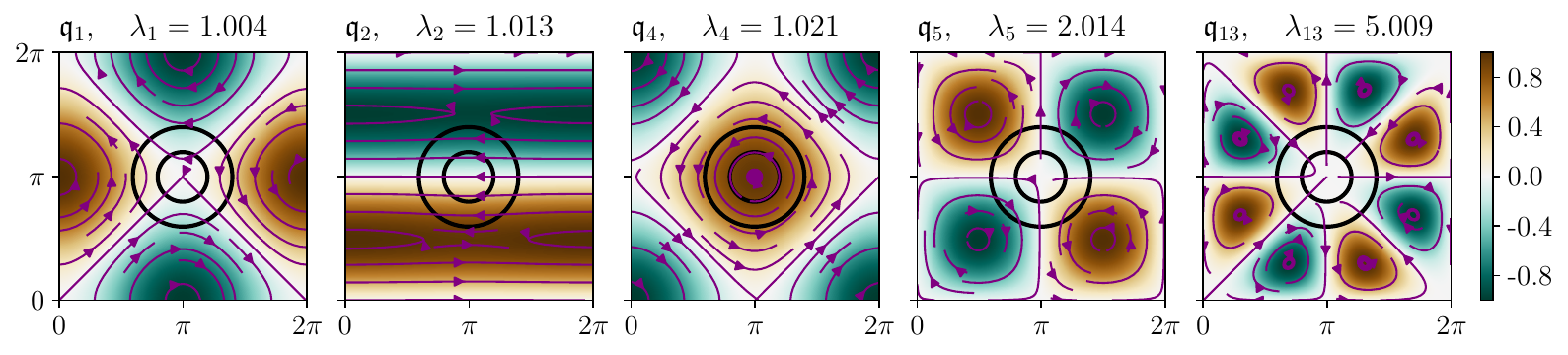}
		\caption{
			Illustrations of the lowest eigenfunctions of the operator \(\mathcal{L}\) with different symmetries.
			The respective eigenvalues are displayed on top of each panel.
			The omitted \(\mathfrak{q}_3\) corresponds to a \(\pi/2\) rotation of \(\mathfrak{q}_2\), as they span the two\-/dimensional eigenspace of the degenerate eigenvalue \(\lambda_2=\lambda_3\).
			Black lines represent the contours at one and two \(\sigma\) of the Gaussian hill.
			Purple arrows point along streamlines of the velocity field.
		}
		\label{fig:eigfn}
	\end{figure*}
	We can interpret this long\-/time state by drawing an analogy with the 2D Navier\-/Stokes and Quasi\-/Geostrophic equations, where the minimum enstrophy principle predicts that the stationary solutions are eigenfunctions of the constitutive relation between \(q\) and \(\psi\) \cite{majda00}.
	To this end, we compute the spectrum of the operator \(\mathcal{L}\) (defined in Eq.~\eqref{eqn:PV_SF}), which consists of eigenvalues of the Laplacian \cite{kuttler84} perturbed by the (small) topography.
	The first eigenfunctions \(\mathfrak{q}_i\) of \(\mathcal{L}\) are sketched in Fig.~\ref{fig:eigfn}.
	While they do not strictly satisfy our forced\-/dissipative equation at stationarity, the flow pattern actually observed in our asymptotic states exhibits a remarkable similarity to the eigenstate corresponding to the smallest non\-/zero eigenvalue, namely \(\mathfrak{q}_1\).
	Since selective decay theory proves that the stationary limit for the aforementioned exemplary fluid equations is their lowest eigenstate, we will refer to the asymptotic configuration of Eq.~\eqref{eqn:fixed_energy} as the ``ground state'' of our system.

	It is important to note that, while the randomness of our initial condition breaks all spatial symmetries, the underlying equations of motion preserve a discrete group of rotations and reflections (see Appendix~\ref{sec:symmetries}).
	Consequently, the basin of attraction for the dynamics is constrained by the symmetry class of the initial data, possibly excluding the previously defined ground state.
	For instance, a rotationally symmetric input would evolve into a flow configuration resembling \(\mathfrak{q}_4\) in Fig.~\ref{fig:eigfn}.
	Overall, taking into account all possible combinations of the conserved symmetries, the dynamics asymptotically converges to one of six flow configurations, each resembling one of the six eigenstates represented in Fig.~\ref{fig:eigfn}.

	Beyond these asymptotic states, the same analogy with minimum enstrophy states can also yield insight into the transient dynamics.
	The metastable state temporarily reached during run (iii) can similarly be associated with an element of the second eigenspace, spanned by \(\mathfrak{q}_2\) and \(\mathfrak{q}_3\).
	This suggests that at high Reynolds numbers ``excited states'' of the system can attain meta\-/stability.
	Accordingly, we expect that, as \(\Rey_L\) is further increased, higher excited states will become metastable, and the dynamics will spend progressively more time trapped in these excited states.
	
	\begin{figure*}[t]
		\centering
		\includegraphics[width=\textwidth]{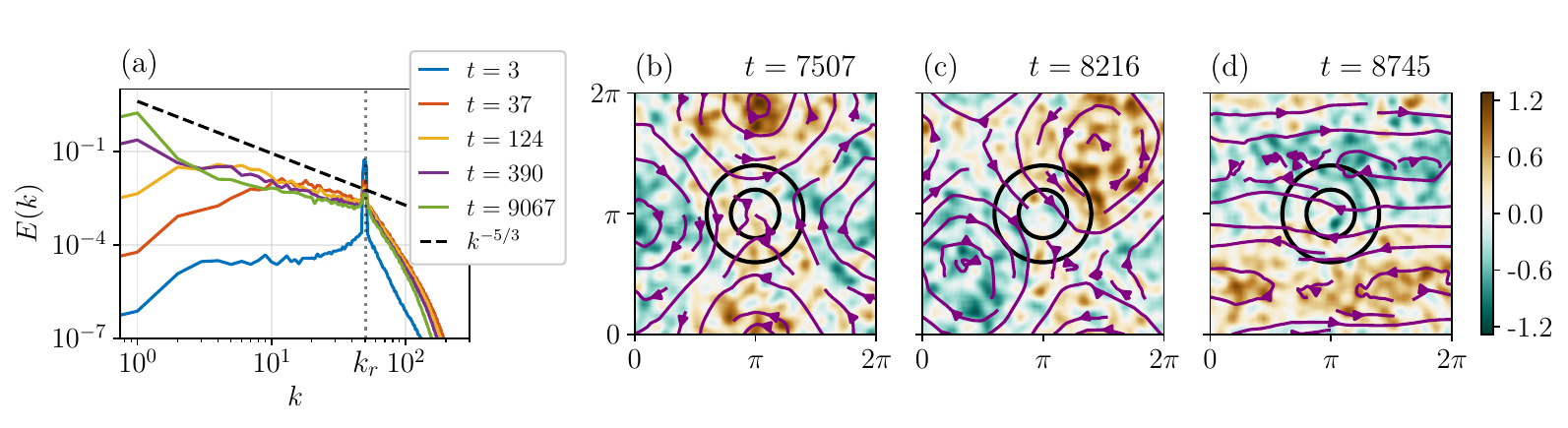}
		\caption{
			Snapshots and energy spectrum of a turbulent flow over a Gaussian hill.
			Panel (a) shows the energy spectrum \(E(k)=\int_{|\vec{k}|=k}\,|\hat{\vec{u}}(\vec{k})|^2/2\,\d\vec{k}\) at different stages of a single time evolution.
			Panels (b, c, d) display representative snapshots of the flow toward the end of the simulation.
			The potential vorticity field, represented in colors, has been smoothened using a Gaussian filter to isolate scales larger than the random forcing.
			Purple arrows point along streamlines of the (non\-/filtered) velocity field.
			Black lines represent the contours at one and two \(\sigma\) of the Gaussian hill.
		}
		\label{fig:turb_plot}
	\end{figure*}
	After isolating the deterministic dynamics of the NRSF model, we now turn our attention to a more realistic scenario where the flow is subjected to an unpredictable external forcing alongside with viscous dissipation.
	A stochastic forcing with a fixed power spectrum can sustain the flow in a turbulent state, compensating the dissipative loss.
	Overall, this setting reproduces, on average, the conditions that allowed for stationary states in the previous section.

	We ran a simulation of Eq.~\eqref{eqn:NRSF_equation} starting from a fluid at rest, \(q_0 = 0\), driven by a stochastic forcing with power input \(\epsilon\simeq2\times10^{-2}\), and viscosity \(\nu=10^{-4}\).
	The topography was again chosen to be a simple Gaussian hill, to compare long\-/time results with the autonomous system.

	The flow evolution exhibits the phenomenology characteristic of 2D turbulence.
	Despite the random forcing averages to zero in time, it overall injects energy into the system, which is redistributed across different modes by the nonlinear advection term.
	Soon an inverse energy cascade builds up (see panel (a) of Fig.~\ref{fig:turb_plot}), leading to the development of large\-/scale structures.
	In the absence of a large\-/scale dissipation mechanism, the energy transported along the inverse cascade accumulates indefinitely, condensing at the lowest mode.
	The simulation is terminated when the kinetic energy reaches a value of \(E\simeq1.5\), corresponding to \(\Rey\simeq1.2\times10^4\), comparable to run (iii) of the deterministic case.

	Panels (b, c, d) of Fig.~\ref{fig:turb_plot} display snapshots of the flow during the late stages of the dynamics.
	The randomly driven flow is observed to self\-/organize into a dipole of domain\-/filling vortices, consistently with previous studies without topographic modulation \cite{smith94}.
	Both vortices avoid the hill, roaming in regions of higher fluid depth \(h\), confirming the behavior observed in the deterministic case.
	However, in this stochastic framework, the concept of a unique ground state is superseded by an ensemble of configurations.
	The three panels show the dipole continuously transitioning between different excited states: the random nature of the forcing prevents the flow from settling in any single configuration.
	Nonetheless, the system remains confined into a region of phase space characterized by a dipole of vortices that consistently avoid the hill.

\section{Conclusions}
	\label{sec:conclusion}
	In this work, we investigated the dynamics of non\-/rotating shallow flows over topography, a regime that has received considerably less attention compared to its rotating counterpart.
	To rigorously address the hydrodynamics of this system, we started from the original 3D Navier\-/Stokes equations and derived a 2D approximation for a 3D incompressible flow. We then developed and validated a numerical scheme based on the mass\-/transport stream function, ensuring the preservation of the equation's symmetries.

	Our central finding extends the standard intuition regarding geophysical flows, demonstrating that in non\-/rotating systems, the nonlinear interaction between flow and bottom topography cannot be neglected.
	Both in controlled deterministic dynamics and in turbulent regimes, long\-/time flow structures consistently organize into maximally spaced vortices that systematically avoid topographic hills, independently of their rotation sign.
	Furthermore, this nonlinear coupling challenges the implications of the selective decay principle on stationarity.
	Unlike quasi\-/geostrophic approximations or pure 2D Navier\-/Stokes, the NRSF model does not necessarily relax toward a unique minimum enstrophy ground state.
	While fluid depth modulation mathematically precludes strict stationarity in purely dissipative system, holding the energy constant reveals that stationary solutions must depend on the Reynolds number.
	Ultimately, at higher Reynolds numbers, the flow can easily be trapped in metastable ``excited'' states, characterized by sub\-/optimal vorticity configurations.

	Further work is needed to better understand the dependence of the long\-/time behavior on the Reynolds number, specifically by exploring the high \(\Rey_L\) limit and seeking analytical stationary solutions to Eq.~\eqref{eqn:stationary_energy}.
	Finally, it would be interesting to examine the influence of more complex topographies on the flow, ranging from the effect of realistic fractal terrains to the inverse cascade blocking over confining topographies.

\appendix

\section{Symmetries of the NRSF equation}
	\label{sec:symmetries}
	While the original Navier\-/Stokes equations preserve a large array of symmetries, constraining the solution and guiding physical intuition, the introduction of a bottom topography may break some of them, requiring further analysis.
	We are here concerned with the spatial symmetries analytically preserved by Eq.~\eqref{eqn:NRSF_equation}.

	A scalar function \(f(\vec{x})\) is symmetric (or anti\-/symmetric) with respect to an isometric endomorphism \(\vec{s}(\vec{x})\) if
	\begin{align}
		\label{eqn:symmetry}
		f(\vec{s}(\vec{x})) = f(\vec{x})\quad\big(\text{or}\,\,-f(\vec{x})\big).
	\end{align}
	We want to understand whether the (anti-)symmetry of a function is preserved by the dynamics, that is, whether the time derivative \(\partial_t{q}\) shares the same symmetry properties of \(q\).
	Eq.~\eqref{eqn:NRSF_equation} includes derivatives, operations involving \(h\), and the alternating Jacobian term.
	The only relevant symmetry operations are the ones compatible with the domain, \(\mathbb{T}\), and shared by the chosen topography \(b\).
	While for a random topography the latter condition excludes any spatial symmetry, for a Gaussian topography we can consider the whole symmetry group of the square, \(D_4\), which consists of rotations by multiples of \(\pi/2\) and reflections about the four axes of symmetry.
	In this case, the column height \(h\), and any operation involving it, are invariant under the action of the group.
	We now examine the symmetry properties of the gradient: applying \(\nabla\) to the symmetry condition \eqref{eqn:symmetry}, using the chain rule and the orthogonality of the Jacobian matrix of an isometry, \(\mathcal{J}_{\vec{s}}(\vec{x})\), one obtains for a (anti-)symmetric function:
	\begin{align}
		\label{eqn:gradient_transform}
		\nabla{f}\big{|}_{\vec{s}(\vec{x})} = \mathcal{J}_{\vec{s}}\,\nabla{f}\big{|}_{\vec{x}}\quad\left(\text{or}\,\,-\mathcal{J}_{\vec{s}}\,\nabla{f}\big{|}_{\vec{x}}\right).
	\end{align}
	Similar computations reveal that the Laplacian commutes with any isometry.
	The Jacobian term \(J(q,\psi)\) depends on the solution \(q\) also through the stream function \(\psi\); given the properties of the Laplacian and of the scalar product of gradients, in the expression of \(\mathcal{L}\) in Eq.~\eqref{eqn:potential_vorticity} and \eqref{eqn:PV_SF}, \(\psi\) and \(q\) share the same symmetry properties.
	Therefore, the way the Jacobian term transforms under the action of \(\vec{s}\) does not depend on whether \(q\) is symmetric or anti\-/symmetric, since \(J\) is effectively quadratic in the solution.
	Nonetheless, because of Eq.~\eqref{eqn:gradient_transform}, the sign of \(J(q,\psi)\) changes in the same way as the unitary skew\-/symmetric matrix \(\mathbb{J}\),
	\[\mathcal{J}_{\vec{s}}^T\mathbb{J}\mathcal{J}_{\vec{s}} = \mathfrak{a}\mathbb{J},\]
	where \(\mathfrak{a} = +1\) for rotations, and \(-1\) for reflections.

	Taking everything into consideration, given a Gaussian topography, the equation of motion only preserves symmetries with respect to rotations by multiples of \(\pi/2\), and anti\-/symmetries with respect to reflections about the symmetry axes of the square.


\begin{thebibliography}{43}
\bibitem [{pressure()}]{footnote:pressure}%
  \BibitemOpen
  \bibinfo {note} {The pressure is completely defined by the hydrostatic condition, \(\partial_z{p} \simeq F_z\), and generally depends on gravity. For our purposes, the actual expression for the pressure is irrelevant, as it will not appear in the final formulation of the vorticity equation of motion.}\BibitemShut {Stop}%
\bibitem [{boundary_terms()}]{footnote:boundary_terms}%
  \BibitemOpen
  \bibinfo {note} {In a domain with a bottom topography, the velocity cannot be horizontal without crossing the boundary, resulting in momentum and energy fluxes through the bottom boundary. The physically accurate procedure would instead consider the vertically\-/averaged velocity \(\langle\vec{u}\rangle_z\), which, in the non\-/dissipative case, yields the same exact equation, through convenient cancellations between the boundary terms (see \cite{zeitlin_book}, chapter 3). On the contrary, repeating a similar derivation for the additional viscous term introduces boundary terms that do not cancel out, \[\Delta\langle\vec{u}\rangle_z+\left[2\nabla\left(\langle\vec{u}\rangle_z-\vec{u}\big{|}_b\right)\cdot\frac{\nabla{h}}{h}+\left(\langle\vec{u}\rangle_z-\vec{u}\big{|}_b\right)\frac{\Delta{h}}{h}\right],\] where \(\vec{u}\big{|}_b\) is the velocity at the bottom. Empirical observation from our long-term integrations shows that these additional boundary terms average out and do not induce any secular drift in the energy, making their net contribution negligible.}\BibitemShut {Stop}%
\bibitem [{jacobian()}]{footnote:jacobian}%
  \BibitemOpen
  \bibinfo {note} {The Jacobian operator is alternating, hence it vanishes whenever stream function and potential vorticity contours are aligned, i.e. \(q = g(\psi)\), for any scalar function \(g(\cdot)\).}\BibitemShut {Stop}%
\bibitem [{stability_test()}]{footnote:stability_test}%
  \BibitemOpen
  \bibinfo {note} {The dynamical stability of the final states has been tested via a finite amplitude perturbation analysis, both with a correlated random noise and with a bred vector \cite{toth93}.}\BibitemShut {Stop}%
\end{thebibliography}
%

\end{document}